\documentclass[10pt,psfig,a4paper,twocolumn]{article}
\usepackage{geometry}
\usepackage{graphics}
\usepackage{setspace}
\usepackage{tabls}
\usepackage[scaled=0.9]{helvet}

\usepackage{color}
\begin{document}
\title{\fontfamily{phv}\selectfont{\huge{\bfseries{
Two remarks about UA5 published data on general characteristics of $p\bar{p}$ collisions 
at $\sqrt{s}$ = 900 GeV.
}}}}
\author{
{\fontfamily{ptm}\selectfont{
\large{\bfseries{M.G. Poghosyan}}\thanks{
Universit\'a  di Torino/INFN, 10125 Torino, Italy.}}
}
}
\date{}
\maketitle
\thispagestyle{empty}
\begin{abstract}
We study UA5 published data on $p\bar{p}$ interaction cross-sections and on charged particles 
pseudorapidity distribution in single-diffractive, non-single diffractive and inelastic events 
and investigate their consistency/inconsistency. 
\end{abstract}
\section{Introduction}
Proton-proton collisions at $\sqrt{s}$ = 900 GeV have already appeared at the LHC and the results 
will be compared with S$p\bar{p}$S data on proton-antiproton collisions at the same center of mass energy 
measured by UA5 collaboration. 
In this paper we point out some nuances concerning UA5 measurements 
and results and show that the uncertainties of measurements are underestimated by UA5. 
This can lead to discrepancies between $pp$ and $p\bar{p}$ at $\sqrt{s}$ = 900 GeV and be wrongly interpreted 
in future theoretical analyses. First we investigate the result of measurements of 
cross-sections and then the result of 
measurements of charged particles pseudorapidity distribution.  
\section{The cross-sections}
In this section we investigate the results of UA5 measuremnt on cross-sections
 of single-diffractive, non-single diffractive and inelastic $p\bar{p}$  interactions.\\
{\bf The measurement of $\sigma_{SD}/\sigma_{NSD}$ ratio at $\sqrt{s} =$ 200 and 900 GeV}.\\
The UA5 detector and event analysis procedures are described in \cite{UA5detector}. Two large
streamer chambers were placed above and below the S$p\bar{p}$S beam pipe. The chambers were triggered by
requiring one or more hits in scintillation counter hodoscopes at each end of the chambers covering
$2 < |\eta| < 5.6$. Two triggers were taken in parallel: a "2-arm" trigger requiring hits at both ends 
to select mainly non single-diffractive events, and a "1-arm" trigger demanding a hit in only one arm
to select highly asymmetric events such as single diffractive events.\\
In case of single-diffraction dissociation UA5 triggered particles produced from anti-proton 
dissociation and  the measured 1-arm triggering cross-section multiplied by factor 2 in order 
to correct for proton dissociation (assuming proton and anti-proton dissociations to be the same).\\
The triggering cross sections $\sigma_1$ and $\sigma_2$ for 1-arm and 2-arm 
triggers are related to the single-diffractive  and non-single diffractive
cross-sections by the trigger efficiencies $\epsilon^{SD, NSD}_{1,2}$ as follows:
\begin{eqnarray}  
\label{Eq:TRXS}
\sigma_1 = \epsilon^{SD}_1\sigma_{SD} + \epsilon^{NSD}_1\sigma_{NSD}, \\
\sigma_2 = \epsilon^{SD}_2\sigma_{SD} + \epsilon^{NSD}_2\sigma_{NSD}. \nonumber
\end{eqnarray}  

Solving Eq. (\ref{Eq:TRXS}) for $\sigma_{SD}/\sigma_{NSD}$ one finds:
\begin{equation}
\label{Eq:XSratio}
\frac{\sigma_{SD}}{\sigma_{NSD}} = \frac{r\cdot\epsilon_{2}^{NSD} - \epsilon_{1}^{NSD}}
{\epsilon_{1}^{SD} - r\cdot\epsilon_{2}^{SD}}.
\end{equation}
Where $r\equiv \sigma_{1}/\sigma_{2}$ and the result of the measurement is \cite{UA5diff}:
\begin{eqnarray*}
r &= & 0.153 \pm 0.015 \,\, at \,\, \sqrt{s} = 200 \, GeV,\nonumber\\
  &= & 0.111 \pm 0.009 \,\, at \,\, \sqrt{s} = 900 \, GeV.\nonumber
\end{eqnarray*}
Determining 1-arm and 2-arm triggering efficiencies for single-diffractive and non-single diffractive 
events based on MC simulations, 
UA5 reported the following value for $R \equiv \sigma_{SD}/\sigma_{NSD}$ \cite{UA5diff}:
\begin{eqnarray}
R &= 0.132 \pm 0.016 \pm 0.024 \,\,\,\,  at \,\,\, \sqrt{s} = 200 \, GeV,\label{Eq:RSdNsd200}\\
  &= 0.180 \pm 0.014 \pm 0.029 \,\,\,\,  at \,\,\, \sqrt{s} = 900 \, GeV.
\label{Eq:RSdNsd}
\end{eqnarray}
The first error is statistical and the second systematic.\\
{\bf The evaluation of triggering efficiencies.}\\
The triggering efficiencies $\epsilon^{SD, NSD}_{1,2}$ in Eq. (\ref{Eq:TRXS}) were estimated by UA5 based 
on Monte Carlo simulations. UA5 detector did not have magnet and was not able to measure  the 
transverse momentum of particles. The MC generator used for simulations was tuned 
to reproduce multiplicity and pseudo-rapidity distribution of particles, but 
simulations were done for different values of mean transverse momentum of particles in order to estimate the 
systematic uncertainties of measurements \cite{UA5diff}.\\
In UA5 MC generator \cite{UA5gener} the cross-section of single-diffraction dissociation as a function of 
diffracted 
system mass  was parametrized as follows:
\begin{equation}
\label{Eq:SDvsM2}
\frac{d\sigma_{SD}}{dM^2} \sim \frac{1}{M^2},
\end{equation}
and masses were generated  in the interval from 1.08 GeV (=$m_{\pi} + m_p$) to $\sqrt{0.05s}$
(see \cite{UA5diff} and \cite{UA5gener} for more detailes).\\
At fragmentation of a diffracted system in single-diffractive interaction the distribution of particles 
is centered around $y_0\simeq\ln(\sqrt{s}/M)$ and covers the rapidity region from
$y_{min} \simeq \ln(\sqrt{s}m_p/M^2)$ to $y_{max} \simeq \ln(\sqrt{s}/m_p)$.
When the mass of the diffracted system is small then the particles are mainly 
concentrated  at the forward region. Increasing the mass of the diffracted system the distribution 
over (pseudo)rapidities moves to mid-rapidities and
the spread of the distribution  becomes wider.
Thus acceptances of different triggers are sensitive to different mass regions of 
diffracted system (at given center of mass energy).
In particular, if the triggers are not placed in very forward region  then the particles produced from 
low-mass diffracted system will not hit the triggers.\\
In Ref. \cite{UA5diff} UA5 claims that masses below 2.5 GeV/$c^2$ were almost never seen by the detector. 
For this purpose they investigated trigger efficiencies enhancing this low-mass region by 
50$\%$ and studying the consequences  of this change in their results.\\
In Table \ref{Tb:TriggEff} we present trigger efficiencies for single-diffractive events as reported by 
UA5 in Ref \cite{UA5diff}. 
\begin{table}[h!]
\begin{center}
\caption{
1-arm and 2-arm trigger efficiencies for single-diffractive events estimated by UA5 in Ref \cite{UA5diff}.
}
\label{Tb:TriggEff}
{
\scriptsize
\begin{tabular}{l l l}
\hline
\, $\sqrt{s} = $200 GeV                 & $\epsilon_{1}^{SD}$  & $\epsilon_{2}^{SD}$ \\
\, $<p_t>$ = 0.55 GeV/c               & $0.61 \pm 0.02$  &  $0.040 \pm 0.004$ \\      
$^{*}\hspace{-1mm}<p_t>$ = 0.45 GeV/c & $0.60 \pm 0.02$  &  $0.048 \pm 0.004$ \\      
\, $<p_t>$ = 0.35 GeV/c               & $0.60 \pm 0.02$  &  $0.051 \pm 0.004$ \\      
\, $<p_t>$ = 0.45 GeV/c    &      &   \\      
\, masses below 2.5 GeV/$c^2$         &  $0.55 \pm 0.02$ & $0.044 \pm 0.004$\\
\, enhanced by 50 $\%$        &                  &\\
\hline
\, $\sqrt{s} = $900 GeV                 & $\epsilon_{1}^{SD}$  & $\epsilon_{2}^{SD}$ \\
\, $<p_t>$ = 0.55 GeV/c               & $0.52 \pm 0.03$  &  $0.10 \pm 0.01$ \\      
$^{*}\hspace{-1mm}<p_t>$ = 0.45 GeV/c & $0.50 \pm 0.01$  &  $0.12 \pm 0.004$ \\      
\, $<p_t>$ = 0.35 GeV/c               & $0.48 \pm 0.02$  &  $0.16 \pm 0.01$ \\      
\, $<p_t>$ = 0.45 GeV/c    &      &   \\      
\, masses below 2.5 GeV/$c^2$         & $0.46 \pm 0.01$ & $0.111 \pm 0.003$\\
\, enhanced by 50 $\%$        &                  &\\
\hline
\end{tabular}
}
\end{center}
\end{table}
Those marked with an asterisk were used for the cross-section calculations, while the others were 
used to calculate the systematic uncertainties.\\
If the triggers of UA5 detector were not sensitive to the masses bellow 2.5 GeV/$c^2$ then the
triggering efficiency for 
the case when this mass region is enhanced by 50$\%$ must be related with the triggering
efficiency marked with an asterisk with the following relation:
\begin{equation}
\frac{\epsilon_{i}^{SD}}{1+\frac{1}{2}\frac{\sigma_{SD}(1.08^2 < M^2< 2.5^2)}{\sigma_{SD}(1.08^2 < M^2< 0.05s)}} 
\,\,\, (i=1,2)
\end{equation}
Using the parameterization (\ref{Eq:SDvsM2}) one can easily evaluate the factor in the denominator:
\begin{equation}
\frac{\sigma_{SD}(1.08^2 < M^2< 2.5^2)}{\sigma_{SD}(1.08^2 < M^2< 0.05s)} = \frac{\ln(2.5^2/1.08^2)}{\ln(0.05s/1.08^2)}.\nonumber
\end{equation}
Thus the "re-normalized" efficiencies will be:
\begin{eqnarray*}
\epsilon^{SD}_{1} = 0.54, & \epsilon^{SD}_{2} = 0.043 & at \,\, \sqrt{s} = 200 \, GeV, \\
\epsilon^{SD}_{1} = 0.46, & \epsilon^{SD}_{2} = 0.0111 & at \,\, \sqrt{s} = 900 \, GeV. 
\end{eqnarray*}
Comparing these numbers with the corresponding numbers in Table \ref{Tb:TriggEff} we 
conclude that at $\sqrt{s}$ = 200 GeV the 
triggers saw some low-mass ($M <$ 2.5 GeV/$c^2$) single-diffractive events but at 
$\sqrt{s}$ = 900 GeV they did not. 
This allows us to claim that at 900 GeV UA5 performed model-dependent extrapolation to 
the low-mass region and the seen cross-section 
of single-diffraction dissociation, $\sigma_{SD}^{HM}$, has multiplied by factor 1.19 
($=\ln(0.05s/1.08^2)/\ln(0.05s/2.5^2)$) 
in order to obtain the "total" single-diffraction cross-section.  Thus in Eq.~(\ref{Eq:RSdNsd}) 
as cross-section of single-diffraction dissociation must be understood the following quantity:
\begin{equation}
\sigma_{SD} = 1.19\cdot\sigma_{SD}^{HM}.
\label{Eq:SDtoHM}
\end{equation}
{\bf Ratios of inelastic interaction cross-sections at 900 and 200 GeV}.\\
In Ref. \cite{UA5inelXS} UA5 reported the result of measurement of the ratio of the inelastic cross-sections 
at $\sqrt{s} =$ 200 and 900 GeV:
\begin{equation}
\label{Eq:Rinel}
R_{inel} \equiv \frac{\sigma_{inel}^{900}}{\sigma_{inel}^{200}} = 1.20 \pm 0.01 \pm 0.02.
\end{equation}
The first error is statistical and the second error systematic which includes contributions 
of background corrections for 1-arm and 2-arm triggers, trigger efficiencies for 
single-diffractive and non-single diffractive processes and luminosity ratio.\\
{\bf The absolute values of the cross-sections.}\\
 In order to obtain the value of inelastic cross-section at 200 GeV UA5 used the following identity:
\begin{equation}
\sigma_{inel}^{200} = \sigma_{tot}^{200} \left[ 1 - \frac{\sigma_{el}^{200}}{ \sigma_{tot}^{200}}
\right]
\end{equation}
The total cross-section at 200 GeV is calculated  based on a fit  
to data on total cross-section from ISR and lower energies \cite{amosNPB262}
which  predicts $\sigma_{tot} = 51.6 \pm 0.4$ mb at $\sqrt{s}$ = 200 GeV.
The value of $\sigma_{el}^{200}/\sigma_{tot}^{200}$ was estimated to be $0.19 \pm 0.01$.
From the paper it is not clear how they estimated this value. 
They just say that they used UA4 measurement  for $\sigma_{el}/\sigma_{tot} = 0.215 \pm 0.005$ at 
$\sqrt{s} = $ 546 GeV \cite{bozzoPLB147} and did an interpolation. 
%
Using the values mentioned above UA5 estimated   
\begin{equation}
\label{Eq:InelXS200}
\sigma_{inel}^{200} = 41.8 \pm 0.6 \, mb,
\end{equation} 
and than using Eq. (\ref{Eq:Rinel}) reported:
\begin{equation}
\label{Eq:InelXS900}
\sigma_{inel}^{900} = 50.3 \pm 0.4 \pm 1.0 \, mb.
\end{equation} 
where the first error is statistical and the second error is
systematical including the error on the estimated values of 
$\sigma_{el}/\sigma_{tot}$ and $\sigma_{tot}^{200}$.\\
Taking into account (\ref{Eq:RSdNsd200}) and (\ref{Eq:RSdNsd}) UA5 obtained the single-diffraction cross-section
\cite{UA5diff}:
\begin{eqnarray}
\sigma_{SD}^{200} = 4.8 \pm 0.5 \pm 0.8, \, mb\\
\sigma_{SD}^{900} = 7.8 \pm 0.5 \pm 1.1. \, mb
\label{Eq:SDXS}
\end{eqnarray}
{\bf What are the consequences of the assumption $\sigma_{SD} = 1.19\cdot\sigma_{SD}^{HM}$ at 900 GeV?}\\
In Table \ref{Tb:sigmainel} we compare UA5 data with predictions of two theoretical 
models \cite{HMD,KMR}. One can see that the result of UA5 on growth of inelastic 
cross-section is slightly smaller from predictions of both theoretical 
models where data from higher energy ($\sqrt{s}$ = 1800 GeV) are used in the fits. 
\begin{table}[h]
\begin{center}
\caption{
Comparisons of both theoretical models on inelastic cross-section with UA5 data.
}
\label{Tb:sigmainel}
\begin{tabular}{c c c c}
\hline
$\sqrt{s}$ GeV  & UA5  Ref. \cite{UA5inelXS} &  Ref. \cite{HMD}  & Ref. \cite{KMR} \\ 
200  &   $41.8 \pm 0.6$              & 41.5  & 43.3 \\ 
900  &   $50.3 \pm 0.4 \pm 1.0 $     & 52.2  &  53.5 \\ 
\hline
\end{tabular}
\label{Tab:table3}
\end{center}
\end{table}
In order to understand this discrepancy, let us remember what is 
measured by UA5 as single-diffractive cross-section and make a detailed comparison 
with the theoretical models. Based on above 
discussions and takeing into account Eq.~(\ref{Eq:SDtoHM}) we write 
(\ref{Eq:RSdNsd}) as follows:
\begin{eqnarray}
\label{Eq:sigmaSDHMNSD1}
\frac{1.19\cdot\sigma_{SD}^{HM,900}}{\sigma_{NSD}^{900}}   = 0.180 \pm 0.014 \pm 0.029,
\end{eqnarray}
and Eq.~(\ref{Eq:Rinel}) as follows:
\begin{equation}
\label{Eq:Rinel1}
\frac{\sigma_{NSD}^{900}+1.19\cdot\sigma_{SD}^{HM,900}}{\sigma_{inel}^{200}} = 1.20 \pm 0.01 \pm 0.02.
\end{equation}
Analogously to Ref. \cite{UA5inelXS}, using (\ref{Eq:InelXS200}) as an input absolute value 
for inelastic cross-section at 200 GeV we obtain:
\begin{eqnarray}
\sigma_{NSD}^{900}  = 42.63 \pm 1.42 \, mb, \\
\sigma_{SD}^{HM,900}= 6.45  \pm 0.92 \, mb.
\end{eqnarray}
These errors include both statistical and systematical errors (added quadratically) of (\ref{Eq:InelXS200}),
(\ref{Eq:sigmaSDHMNSD1}) and (\ref{Eq:Rinel1}).\\
In Table \ref{Tb:comp} we compare the predictions of both theoretical models with data from UA5. 
\begin{table}[h!]
\begin{center}
\caption{
Comparison of predictions of two theoretical models with UA5 data. 
}
\label{Tb:comp}
{
\scriptsize
\begin{tabular}{l l l l}
value & UA5   &  Ref.\cite{HMD}  & Ref.\cite{KMR} \\
\hline
$\sigma_{SD}^{HM,900}$                  & 6.45 $\pm$ 0.92              & 6.4    &  5.6  \\[.05cm]
$\sigma_{SD}^{LM,900}$                  & 1.23 $\pm$ 0.17              & 2.9    &  3.9  \\[.05cm]
$\sigma_{NSD}^{900}$                    & 42.63 $\pm$ 1.42             & 42.9   &  44  \\[.05cm]
$\sigma_{SD}^{HM,900}/\sigma_{NSD}^{900}$ & 1.51 $\pm$ 0.012 $\pm$ 0.024 & 0.149 & 0.127 \\[.05cm]
$\sigma_{SD}^{900}/\sigma_{NSD}^{900}$    & 1.80 $\pm$ 0.014 $\pm$ 0.029 & 2.17 & 2.16 \\[.05cm]
$\sigma_{inel}^{900}/\sigma_{inel}^{200}$ & 1.20 $\pm$ 0.01  $\pm$ 0.02  & 1.26  & 1.24  \\[.05cm]
$\frac{\sigma_{NSD}^{900} + 1.19 \cdot \sigma_{SD}^{HM,900}}{\sigma_{inel}^{200}}$
                                       & 1.20 $\pm$ 0.01  $\pm$ 0.02  & 1.21  & 1.27  \\[.05cm]
\end{tabular}
}
\end{center}
\end{table}
One can see the predictions of both theoretical models are in good agreement with the UA5 data which 
are "really" measured and deviations appear from the extrapolation of single-diffraction 
cross-section to low-mass region.\\
%
 In Regge theory (see \cite{KaidalovPR} for details) the $\sim 1/M^2$ dependence 
can be obtained assuming the intercept of the Pomeron to be unity in the cross-section 
corresponding to the triple-Pomeron diagram. Nevertheless, it is a known fact (and it was 
known at the times of UA5) that the intercept of the Pomeron is bigger than one and the 
cross-section falls steeper ($\sim 1/M^{2(1+\Delta)}$, $\Delta >0 $). In addition to this, 
at low-mass single-diffraction the contribution of $PPR$ vertex is essential 
($\sim 1/M^{2(1.5+2\Delta)}$) which falls much steeper than the former one. Thus the low-mass 
single-diffraction cross-section is bigger than the one expected from $1/M^2$ dependence 
if one fixes the spectra at high-masses. This is the reason of discrepancy between the 
results of extrapolation of both theoretical models and UA5.\\
It must be stressed that the low-mass single-diffraction is experimentally studied in 
details up to at ISR energies (a theoretical analysis of these data can be found 
in \cite{KaidalovPR}). From S$p\bar{p}$S and Tevaton experiments we do not know much about it.
 Nevertheless, UA4 \cite{UA4} from S$p\bar{p}$S reported that at $\sqrt{s}$ = 546 GeV the 
measured cross-section of single-diffraction in the region of masses $M < 4$ GeV/$c^2$ is 
higher by about a factor two than the one expected from the extrapolation with $1/M^2$ 
dependence from high-masses to low-masses. This experimental fact argues in favor of 
both theoretical models used in this analysis.\\
So, the value or the systematic error (of amount 1 mb, which should include the uncertainty 
of extrapolation to the single-diffraction dissociation cross-section to the low-mass region) 
assigned by UA5 to the inelastic and single-diffractive cross-sections at 900 GeV in (\ref{Eq:InelXS900}) and
(\ref{Eq:SDXS})
is underestimated  by at least $2\div3$ mb.
\section{Pseudorapidity distribution}
\begin{figure}[h!]
\begin{center}
\resizebox{0.4\textwidth}{!}{\includegraphics{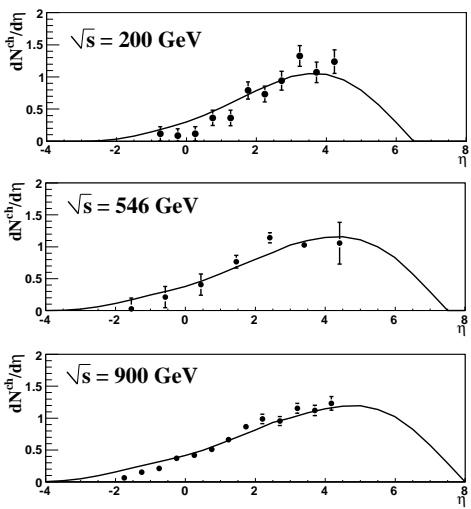}}
\caption{Comparison of QGSM predictions with UA5 data on charged particles pseudorapidity distribution in SD events.}
\label{Fig:dNdEtaSD}
\end{center}
\end{figure}

\begin{figure}[h!]
\begin{center}
\resizebox{0.4\textwidth}{!}{\includegraphics{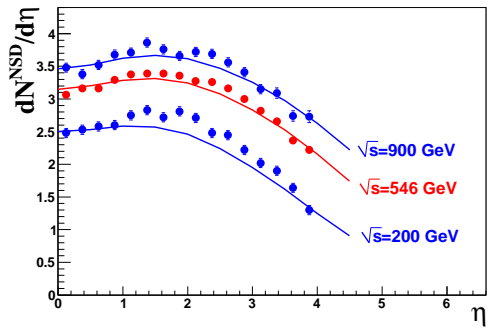}}
\caption{Comparison of QGSM predictions with UA5 data for charged particles pseudorapidity distribution in NSD events.}
\label{Fig:dNdEtaNSD}
\end{center}
\end{figure}

\begin{figure}[h!]
\begin{center}
\resizebox{0.4\textwidth}{!}{\includegraphics{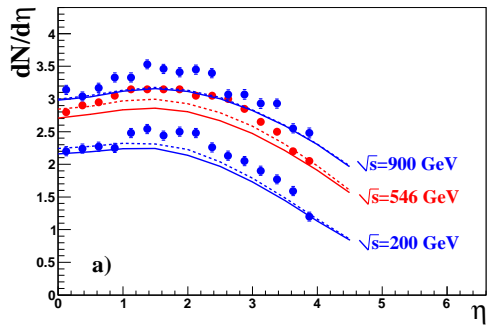}}
\caption{
Comparison of QGSM predictions with UA5 data for charged particles pseudorapidity distribution in inelastic events. 
}
\label{Fig:dNdEtaInel}
\end{center}
\end{figure}

In this section we consider the results of UA5 measuremnt on
charged particles pseudorapidity distribution in single-diffractive, non-single diffractive and 
inelastic $p\bar{p}$ interactions  and discuss their consistency/inconsistency.\\
Although the UA5 collaboration in \cite{UA5diff} studied only diffraction dissociation of proton, 
the final results for inelastic events are expected to be corrected also for anti-proton dissociation.
We compare the predictions of Quark-Gluon String Model (QGSM, see \cite{QGSM, QGSM1}) for 
single and non-single diffractive events with the data 
and later for inelastic ones. In order to calculate the pseudorapidity distribution of charged particles 
for inelastic events we mixed single-diffractive and non-single diffractive events with the weights 
predicted by the theoretical model as well as with the weights deduced 
from experimental data.\\
{\bf Charged particles density $dN_{ch}/d\eta$ as a function of $\eta$ in single-diffractive and non-single 
diffractive events}\\
QGSM predictions on charged particles pseudorapidity in single-diffractive and non-single 
diffractive events are in reasonable agreement with 
UA5 data (see Figs \ref{Fig:dNdEtaSD} and \ref{Fig:dNdEtaNSD}).
UA5 data are taken from \cite{UA5diff, UA5InelNSD, UA5PRP} and the details of theoretical calculations can 
be found in \cite{QGSM1}.
We stress that the description of these data is achieved in a parameter-free way. 
\newline
{\bf Charged particles density $dN_{ch}/d\eta$ as a function of $\eta$ in inelastic events}\\
By definition, in inelastic collisions of two hadrons the diffractive dissociations of each of the 
incoming hadrons may take place separately (single-diffractive dissociation), and in addition the 
non-single diffractive interactions may also take place. This is illustrated by (\ref{Eq:Inel}) for proton-antiproton 
collision.
\begin{equation}
p + \bar{p} \rightarrow \underbrace{ pX_1 + X_2\bar{p}}_{SD} + NSD
\label{Eq:Inel}
\end{equation}
In Fig. \ref{Fig:dNdEtaInel} we compare the predictions of QGSM with the experimental data from UA5 
(\cite{UA5InelNSD, UA5PRP}) for pseudorapidity distribution of charged particles in inelastic events. In order 
to calculate the $dN_{ch}/d\eta$ for inelastic events we summed charged particles pseudorapidity 
distributions of single-diffractive and non-single diffractive events with the default theoretically 
calculated weights (solid lines) 
\cite{HMD} as well as with the weights deduced from the experimental data (dotted lines) \cite{UA5diff, UA5PRP}.
We calculate $dN/d\eta$ using the following expression:
\begin{equation}
\frac{dN}{d\eta} = \frac{\sigma^{NSD}\frac{dN^{NSD}}{d\eta}+\frac{\sigma^{SD}}{2}\frac{dN^{p\bar{p}\rightarrow pX}}{d\eta}
+\frac{\sigma^{SD}}{2}\frac{dN^{p\bar{p}\rightarrow X\bar{p}}}{d\eta}}{\sigma^{NSD}+\sigma^{SD}}.
\end{equation}
As we see, the predictions of QGSM for inelastic evens are significantly different from experimental data in 
$\eta$ region from 1 to 3, 
whereas the predictions for single-diffractive and non-single diffractive events are described well enough.\\

\section{Summary}
At $\sqrt{s}$ = 900 GeV $p\bar{p}$ single-diffractive  interactions UA5 triggers 
were not able to register 
particles produced from diffracted systems with masses below 2.5 GeV/$c^2$. For correcting 
inelastic and single-diffractive cross-sections for this low-mass diffraction region UA5 
used $d\sigma/dM^2 \sim 1/M^2$ simple parameterization which resulted to underestimation 
of inelastic and single-diffraction dissociation cross-sections by $2\div3$ mb.\\
From an analysis of UA5 data for charged particles pseudorapidity distributions in single-diffractive, 
non-single diffractive and inelastic events we conclude that UA5 data are not self-consistent.\\
These factes must be taken into account during the tuning 
of theoretical models (and MC generators) and at further comparison of LHC $pp$ data with UA5 $p\bar{p}$ data. 

\section{Acknowledgements}
I thank A.Kaidalov, V.Khoze, K. Safarik, J. Schukraft and J.-P. Revol for  discussions. 
%

\end{document}